\documentstyle[sprocl,epsfig]{article}
\begin{document}
\title{Summary of RADCOR 98}
\author{R. D. Peccei}
\address{Department of Physics and Astronomy, UCLA, Los Angeles, 
CA 90095-1547}
\maketitle
\begin{abstract}
This summary is organized in four parts.  In the first part results
in the electroweak theory are discussed, including precision
tests of the Standard Model.  The second part deals with recent
results in QCD, focusing on areas where meaningful comparisons
between theory and experiment are possible.  The third part
summarizes some of the salient technical progress in studying
two-loop radiative effects in a variety of contexts, as well as
progress made in calculating
radiative corrections in the LEP 200 region.  Finally, in the fourth
part, a discussion of the effects of radiative corrections, both
as a result of new physics and in new energy regimes, is presented
focusing on their future experimental implications.
\end{abstract}

\section{Introduction}

This Fourth Symposium on Radiative Corrections, following earlier meetings at
Brighton, Knoxville and Krakow, amply demonstrated the vitality and
relevance of this topic.  Radiative corrections are the bridge that joins
quantum field theory to phenomenology.  In the electroweak theory they help
validate the Standard Model and circumscribe possible new phenomena.  At
Barcelona, there were strong hints that the ``old" success of radiative
corrections in predicting the top mass may well be followed by a new success,
that of predicting a light Higgs.  In more established theories, like QED and
QCD, radiative corrections help to organize the way one thinks about
phenomena which by themselves are non-perturbative in nature---phenomena like
confinement, hadronization and bound state decays.

There were far too many talks in RADCOR 98 to be able to summarize them in
detail, even if I could!  Hence, I opted instead to make 
comments on some of the results and give my overall impressions in different areas.  Roughly speaking,
these comments can be organized under four rubrics:  Results in the Electroweak Theory;
QCD Results; Technical Progress; and Future Tests.  The above divisions,
however, are not sharp.  For instance, QCD radiative corrections are clearly
important for precision electroweak tests.  Also, much of the technical progress
involves re-expanding or resumming certain QCD and/or electroweak results.
Finally, future tests often involve radiative effects in new regimes of the
Standard Model itself.  In a nutshell, one could summarize the conclusions for
these four topics very simply:  all is well in the electroweak sector; QCD
works; impressive technical advances continue to be made; and future
experiments will surely open up new physics windows.  However, let me be a 
bit more specific!

\section{Results in the Electroweak Theory}

Substantial new data has sharpened our understanding of the electroweak
theory and its parameters.  At RADCOR 98 the experimental situation was ably
summarized in three separate talks.  F. Teubert~\cite{Teubert} reviewed
precision tests of the standard model at the $Z$ resonance; E. Lan\c{c}on~\cite{Lancon} talked about LEP 200 results, particularly those concerning
$W$ physics; and M. Tuts~\cite{Tuts} discussed results on top 
and $W$'s coming
from the Tevatron Collider.  Wolfgang Hollik,\cite{Hollik} in his own
overview talk, discussed how this data compared in detail with the standard
electroweak theory.  He concluded that there was total consistency between
theory and experiment at the level of accuracy of a tenth of a percent---a
remarkable feat!
Let me briefly outline the main results presented.

\subsection{Top}

The CDF and DO combined results for the top mass, presented at this
conference by Tuts,\cite{Tuts} now make top the quark whose mass is best
known.  The combined result
\begin{equation}
m_t = (173.8 \pm 3.2 \pm 3.9)~{\rm GeV} =
(173.8 \pm 5.0)~{\rm GeV}
\end{equation}
has a relative error $\delta m_t/m_t$ of less than 3\%---an extraordinary result.
In addition, CDF and DO have a quite accurate determination of the top pair
production cross-section~\cite{Tuts}
\begin{equation}
\sigma_{t\bar t} = \left\{ \begin{array}{ll}
(5.9 \pm 1.7)~{\rm pb} & {\rm DO} \\
\left(7.6 ^{\textstyle + 1.8}_{\textstyle -1.5} \right)~{\rm pb} & {\rm CDF}
\end{array}
\right.
\end{equation}
in good agreement with the theoretical QCD predictions, which range from
4.7 to 6.2 pb.

\subsection{$W^\pm$}

Precise values for $M_W$ are inferred from studies of the process
$e^+e^-\to W^+W^-$ at LEP2~\cite{Lancon} and from $W$ production at the Tevatron.\cite{Tuts}  Combining the threshold analysis of the $W$ mass at
$\sqrt{s} = 161~{\rm GeV}$ with the value of $M_W$ obtained by direct
reconstruction at both $\sqrt{s} = 172~{\rm GeV}$ and $\sqrt{s} = 183~{\rm GeV}$, the averaged results from the four LEP collaborations determine the
$W$ mass to 90 MeV~\cite{Lancon}
\begin{eqnarray}
M_W &=& (80.37 \pm 0.07 \pm 0.04 \pm 0.02)~{\rm GeV} \\
&=& (80.37 \pm 0.09)~{\rm GeV}~.
\end{eqnarray}
In the above, the dominant error (70 MeV) is statistical, with about 40 MeV
coming from not being able to disentangle final state interactions between
the two produced $W$'s and 20 MeV arising from uncertainties in the beam
energy.

A similar error is obtained by combining the values for $M_W$ obtained by the
CDF and DO collaborations (as well as 
from the old UA2 data),\cite{Tuts} giving a
``collider value" for $M_W$ of
\begin{equation}
M_W = (80.40 \pm 0.09)~{\rm GeV}~.
\end{equation}
The world average for $M_W$, determined from the above two direct measurements,
has an error of 60 MeV~\cite{Karlen}
\begin{equation}
M_W|_{\rm direct} = (80.39 \pm 0.06)~{\rm GeV}~,
\end{equation}
so that now we know the $W$ mass to better than 1 part per mil.  This value
is in good agreement with 
the new NUTEV result from deep inelastic neutrino scattering~\cite{NUTEV}
for the weak mixing angle
\begin{equation}
(\sin^2\theta_W)_S = 1- M^2_W/M^2_Z =
0.2253 \pm 0.0019 \pm 0.0010~,
\end{equation}
which determines the $W$-mass to an accuracy of 110 MeV
\begin{equation}
M_W = (80.26 \pm 0.11)~{\rm GeV}~.
\end{equation}
Eq. (6) also agrees very well with the
very precise indirect $W$ mass determination obtained from a global fit of all
other high precision eletroweak data, which gives~\cite{Teubert}
\begin{equation}
M_W|_{\rm indirect} = (80.365 \pm 0.030)~{\rm GeV}~.
\end{equation}
I comment below on this latter fit and its implications.

\subsection{Precision Tests at the $Z$-Resonance.}

Precision measurements at the $Z$ resonance,\cite{Teubert} plus a knowledge
of $m_t$ and $M_W$, overconstrain the Standard Model.  Thus, as Hollik~\cite{Hollik} emphasized, present-day data provides rather significant
tests of the electroweak theory.  Fits of all electroweak data to the
Standard Model are in terrific agreement with expectations, with very few
quantities in the fit being over $2\sigma$ away from the fit value.  This is
nicely seen in Fig. 1 which summarizes the Standard Model analysis presented
by Gr\"unewald~\cite{Gruenwald} at the International Conference on High
Energy Physics in Vancouver this summer.  Not only is the data consistent with
the Standard Model, but as Teubert~\cite{Teubert} 
emphasized it is also internally consistent.
This was most clearly seen in the comparison of different determinations of
$\sin^2\theta_W^{\rm eff}$ at both LEP and SLD which are also at most $2\sigma$ away from
the average value
\begin{equation}
\sin^2\theta^{\rm eff}_W = 0.23155\pm 0.00018~.
\end{equation}

\begin{figure}
\center
%\rule{2cm}{0.2mm}\hfill \rule{2cm}{0.2mm}
%\vskip 4cm
%\rule{2cm}{0.2mm}\hfill \rule{2cm}{0.2mm}
\epsfig{file=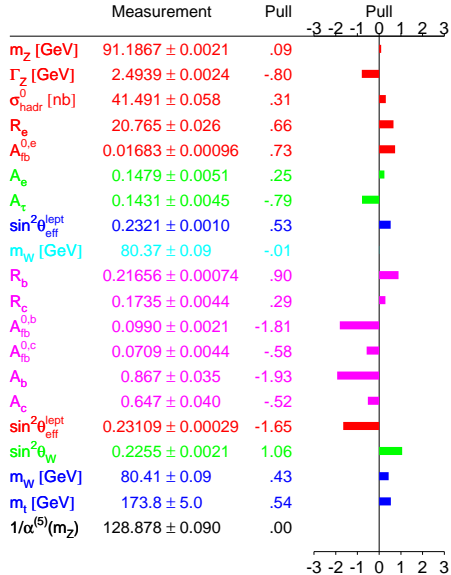,height=3in}
\caption{Standard Model fit.}

\end{figure}

In the Standard Model, given $G_F,~\alpha,~M_Z$ and $m_t$, the only free
parameter remaining is the Higgs mass $M_H$.  Unfortunately, even the present
high precision data does not give a strong constraint on $M_H$, since the
effects of the Higgs mass are only proportional to $\alpha \ln M_H$.
Nevertheless, the 68\% CL contours in the $M_W-m_t$ plane shown in Fig. 2,
determined both through Standard Model fits and by the direct measurements
of $m_t$ and $M_W$, favor a low value for the Higgs mass
\begin{equation}
M_H|_{\rm rad.~corr.} = \left(84^{\textstyle +91}_{\textstyle -51}\right)~
{\rm GeV}~,
\end{equation}
leading to a one-sided 95\% CL bound for $M_H$ of $M_H < 262~ {\rm GeV}$.\cite{Teubert}

\begin{figure}
\center
%\rule{2cm}{0.2mm}\hfill \rule{2cm}{0.2mm}
%\vskip 4cm
%\rule{2cm}{0.2mm}\hfill \rule{2cm}{0.2mm}
\epsfig{file=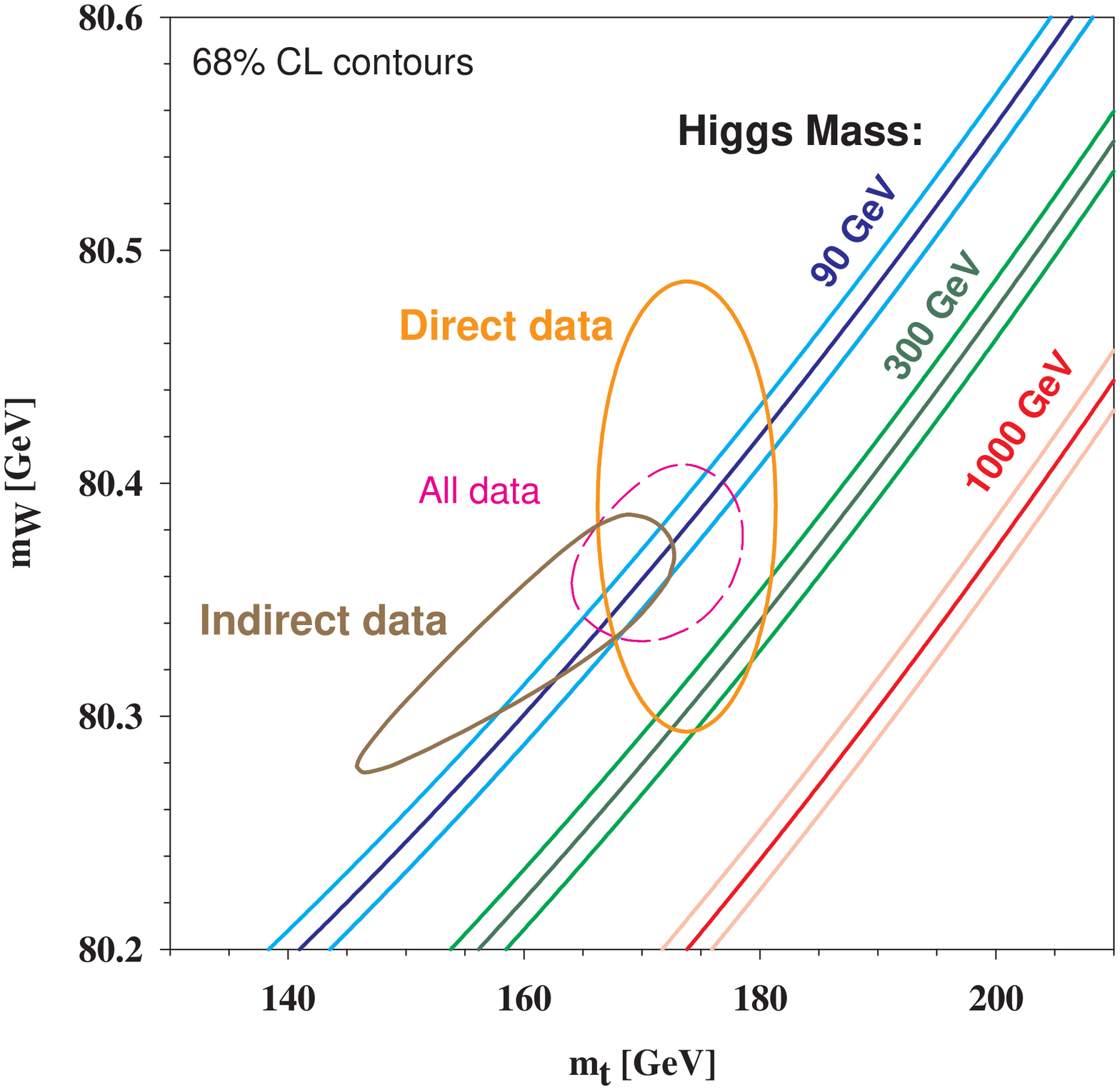,height=3in}
\caption{Allowed contours in the $M_W$- $m_t$ plane.}
\label{fig:radk}
\end{figure}

Errors in $G_F$ and $M_Z$ are insignificant for 
precision tests, while errors
in $m_t$ and $\alpha(M_Z^2)$, as well as the $M_H$ dependence, affect
different quantities differently.  For instance,\cite{Teubert} $\sin^2\theta^{\rm eff}_W$ is most strongly dependent on the error in
$\alpha(M_Z^2)$ and on the Higgs mass dependence, while~\cite{Gambino}
$R_b$ is most sensitive to $\delta m_t$.  Because the radiative effects
have a quadratic sensitivity on $m_t$, precision electroweak data alone 
provide a strong ``prediction" for the top mass.  Indeed, if $m_t$ is taken
as unknown in the fit, then one predicts~\cite{Teubert}
\begin{equation}
m_t|_{\rm rad.~corr.} = \left(161.1 ^{\textstyle +8.2}_{\textstyle -7.1}\right)~
\rm GeV~.
\end{equation}

\subsection{Higgs}

Fig. 3 shows the present situation regarding the Higgs mass.  The vertical
line displays the direct mass limit on the Higgs obtained at LEP 200~\cite{Lancon} from Higgs searches in the process $e^+e^-\to ZH$.  Using the
$\sqrt{s} = 183~{\rm GeV}$ data for all four LEP experiments~\cite{McNamara}
one finds:
\begin{equation}
M_H > 89.8~{\rm GeV} ~~~~~~~ (95\% C.L.)
\end{equation}
I will discuss a little later on the error band on the Higgs mass in Fig. 3.  
However, here
I want to discuss briefly the prospect of ``improving" the Higgs
upper bound by reducing the error in $\alpha^{-1}(M_Z^2)$.  As can be seen in
Fig. 3, the changes effected by using instead of the ``standard value"
$\alpha^{-1}(M_Z^2) = 128.878 \pm 0.090$, the ``improved value" 
$\alpha^{-1}(M_Z^2) = 128.905 \pm 0.036$ are quite significant.  As Jegerlehner~\cite{Jegerlehner} pointed out at RADCOR 98, the improvement in the
error is essentially due to the perturbative QCD improved value for
$\delta\Delta \alpha^5$, which is reduced by about a factor of 5.  Jegerlehner
thought this reduction 
to be highly optimistic, not because the theoretical method is
unreliable but because it is difficult to identify the experimental 
contributions
to this accuracy.  In addition, as Gambino~\cite{Gambino} emphasized, real
improvements in the Higgs upper bound need a parallel reduction in $\delta m_t$
and $\delta M_W$ to about 2 GeV and 30 MeV, respectively.

\begin{figure}[t]
\center
%\rule{2cm}{0.2mm}\hfill \rule{2cm}{0.2mm}
%\vskip 4cm
%\rule{2cm}{0.2mm}\hfill \rule{2cm}{0.2mm}
\epsfig{file=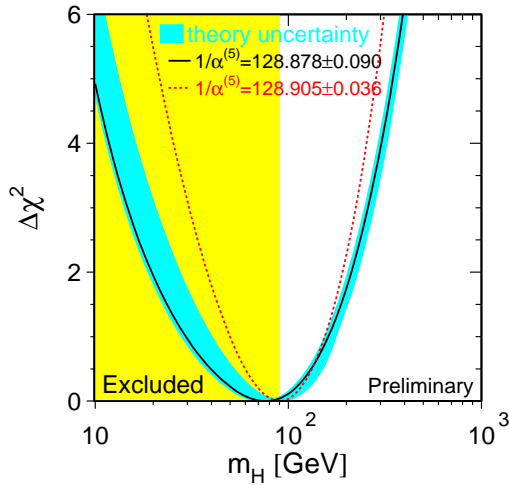,height=2.5in}
\caption{Allowed region for $M_H$ from radiative corrections.}
\label{fig:radk1}
\end{figure}

\subsection{Theoretical Remarks.}

There are two theoretical 
remarks on precision tests, made at RADCOR 98,
which are worth repeating here.  It is often said that the precision tests
of the electroweak theory essentially are only sensitive to the running of
$\alpha$ from $\alpha(0)$ to $\alpha(M_Z^2)$.  However, as Hollik explained,~\cite{Hollik} this is not really the case.  
For instance, one now has a value
for the $\rho$ parameter which is different from unity at $3\sigma$:
\begin{equation}
\rho = 1.0042 \pm 0.0012~.
\end{equation}
Also, when one examines the radiative shift $\Delta r$, it is not true that
the result is dominated by the running of $\alpha$.  Rather,  schematically,~\cite{Hollik} one has
\begin{equation}
\Delta r = \Delta\alpha - (\cos^2\theta_W/\sin^2\theta_W)
\Delta\rho + \Delta r_{\rm rest}~.
\end{equation}
While the first term contributes at the 6\% level, the others contributes at
the 4\% level and 1\% level, respectively. So, really, $\Delta\alpha$ does not
dominate.

A second important point to make~\cite{Hollik,deBoer} is that an
excellent Standard Model fit to the data {\bf does not} exclude an equally
good MSSM fit.  Rather, the good agreement of the data with the Standard
Model serves to provide useful constraints for sparticles and SUSY 
parameters, but does not exclude having supersymmetry at or near, the weak
scale.  For example,   
de Boer,\cite{deBoer} showed a very nice fit of all the
precision data (of similar quality to that of Fig. 1) in the MSSM, provided
one took $|\mu| > m_{1/2}$; $m_{\tilde q} \sim 300~{\rm GeV}$; 
$m_h < 105~{\rm GeV}$
and $\tan\beta = 1.65 \pm 0.3$.  While specific results are model dependent,
two points appear common in these MSSM fits.  First, as is well known and as
I will discuss more later on, in supersymmetric theories 
there is always a light Higgs.  Second, however, the light
Higgs in the MSSM gives rise to a band which is systematically above the
Standard Model light Higgs band (see Fig. 2) in the $M_W-m_t$ plane.
From this point of view, decreasing the errors in the top and $W$ masses would be very important, as it could discriminate between these two 
alternatives.~\cite{Hollik}

\section{QCD Results}

In his plenary talk on QCD at the International Conference on High Energy
Physics in Vancouver this summer, Yuri Dokshitzer\cite{Dok} made the remark
that since QCD is a mature theory ``the issue is not to check QCD, but to
see how it works".  At RADCOR 98 many beautiful examples were presented of
how indeed QCD works.  I will discuss some of them here.

\subsection {$b\to s\gamma$}

Although the process $b\to s\gamma$ is an electroweak process, it is
particularly sensitive to QCD corrections since these corrections change the
nature of the GIM cancellation from quadratic $[m_t^2-m_c^2]$ to
logarithmic $[\ln m_t^2/m_c^2]$.  Gambino~\cite{Gambino} discussed the NLO
analysis of this process in Barcelona.  The result obtained for the branching
ratio for this process in NLO
\begin{equation}
\left.{\rm BR}(B\to X_s\gamma)\right|_{\rm NLO} =
(3.29 \pm 0.30) \times 10^{-4}
\end{equation}
is considerably larger than the lowest order result:
\begin{equation}
\left.{\rm BR}(B \to X_s\gamma)\right|_{\rm LO} =
(2.46 \pm 0.72) \times 10^{-4
}~.
\end{equation}
More importantly, this result is in excellent agreement with the new
CLEO (and ALEPH) 
branching ratio discussed at the meeting by G. Eigen~\cite{Eigen}
\begin{equation}
\left.{\rm BR}(B\to X_s\gamma)\right|_{\rm CLEO} =
(3.15 \pm 0.35 \pm 0.32 \pm 0.26)\times 10^{-4}~,
\end{equation}
where the last error above is an estimate of the uncertainty caused by the
models used to extract this branching ratio from the data.

\subsection{Deep Inelastic Scattering.}

At RADCOR 98 there were two presentations~\cite{Hernandez,Lahmann} of the
most recent results from HERA.  These talks demonstrated in a beautiful way
QCD at work in deep inelastic scattering.  This was apparent in the wonderful and very precise data on $F_2(x,Q^2)$, where one could see with the naked
eye very little $Q^2$-dependence for large $x$ and considerable dependence at low $x$--just what QCD ordered!  As 
Forte~\cite{Forte} emphasized in his talk, the scaling violations at small $x$
can be used to extract simultaneously the gluon distribution and
$\alpha_s(Q^2)$ since, in this region, approximately,
\begin{equation}
\frac{dF_2}{d\ln Q^2} \sim \alpha_s(Q^2) xg(x,Q^2)~.
\end{equation}
This analysis is apparently under way, but has not been done yet.  What has
been done, however, is to get a first idea of the gluon distribution function
from the behavior of the di-jet cross sections at HERA.~\cite{Lahmann}  This
distribution function, as expected, rises sharply as $x$ becomes smaller.
However, the errors are still too large to effectively discriminate among
different pdf parametrizations.

Hernandez~\cite{Hernandez} discussing the sharp rise at low $x$, for fixed
$Q^2$, of $F_2(x,Q^2)$, suggested that possibly 
this rise was not accountable
by the usual DGLAP evolution equations.  Forte,\cite{Forte} however, was
not convinced.  Indeed, he discussed the analysis tht he and Ball~\cite{BF} did
of earlier HERA data on $F_2(x;Q^2)$ at small $x$ (and for a range of $Q^2$),
where using DGLAP they actually were able to extract already quite a good
value for $\alpha_s(M_Z^2)$:
\begin{equation}
\alpha_s(M_z^2)|_{\rm BF} = 0.120 \pm 0.005 \pm 0.009~.
\end{equation}
Clearly, as discussed above, one awaits a full analysis of the evolution of
$F_2(x,Q^2)$ at small $x$ by the HERA experimental 
collaborations to get more
accurate results on both $\alpha_s$ and the gluon distribution function.

\subsection{Jet Physics.}

Hadronic jets are ubiquitous features of hard scattering processes and are
well described by QCD.  However, for more detailed observables, comparison
of data with QCD is more subtle.  A nice example of the subtleties involved
was discussed by Bethke~\cite{Bethke} at RADCOR 98, involving the ratio of
multiplicities of gluon jets relative to quark jets.  In QCD, as is well
known, this ratio at asymptotic energies is simply given by a color factor, with
$\langle n\rangle_g/\langle n\rangle_q = 9/4$.  One can  nicely study gluon jets at LEP by focusing on $Z\to b\bar bg$, since one can identify the $b$-jets.
If one does this with no selection in the data, one finds a, dissapointingly
low, result:
\begin{equation}
\left.\frac{\langle n\rangle_g}{\langle n\rangle_q}\right|_{\rm no-selection}
= 1.27 \pm 0.07~.
\end{equation}
On the other hand, if one selects events in which the $b\bar b$ jets are
almost collinear and back to back to the gluon jets---thus producing gluon jets
of rather high energy $(\langle E\rangle_g = 42~{\rm GeV})$---one obtains
a much higher number
\begin{equation}
\left.\frac{\langle n\rangle_g}{\langle n\rangle_q}\right|_{\rm selected} =
1.87 \pm 0.05 \pm 0.12~.
\end{equation}
The conclusion to be drawn is that, even for these selected events, the
energies involved are not yet sufficient to expect the asymptotic QCD
prediction to hold.  Nevertheless, with these cuts there is a marked shift
of the results towards the QCD expectations.

Bethke~\cite{Bethke} also showed that NLO QCD (with some resummation)
describes well three- and four-jet distributions at LEP 100, giving tight
errors on $\alpha_s(M_Z^2)$ and a mild scale dependence.
One can also look at event shape variables obtaining fits where power
corrections are correlated, as one would expect from renormalons.\cite{Zakharov}
For instance, one finds 
\begin{equation}
\frac{1}{2} \langle 1-T\rangle = \frac{1}{3\pi} \langle C\rangle~.
\end{equation}

Fits with optimized scales give even smaller errors in $\alpha_s$ for shape
variables,\cite{Bethke} but here there is some controversy.  For
example, DELPHI has done an analysis of 18 shape variables and extracted
$\alpha_s$ by a simultaneous fit in which also the scale $\mu^2$ was optimized
for each variable.\cite{Duchesnau}  The resulting value of $\alpha_s$ has a very small error
\begin{equation}
\alpha_s(M_Z^2) = 0.1164 \pm 0.0025,
\end{equation}
but there is considerable variation in the scale parameter, ranging from
$\mu^2/M_Z^2 = 0.0033$ to $\mu^2/M_Z^2 = 6.33$.  This seems an enormous range
to me!
For this reason, it seems much more sensible instead, as advocated
emphatically by Brodsky~\cite{Brodsky} at this meeting, to use a fixed 
but {\bf physical} scale to do the analysis of jet and shape variables.
Brodsky rather naturally favors the BLM scale~\cite{BLM} in which different
processes are related through appropriate rescalings.  Thus $\alpha_s$ for
a given process is related to $\alpha_s$ in another process through a rescaling,
such that in the resulting 
power series expansion the coefficients are the conformal
coefficients:
\begin{equation}
\alpha_s(Q_A^2) = \alpha_s(Q^2_B)
\left[1 + r_1^{B/A} \frac{\alpha_s}{\pi} + ... \right],
\end{equation}
with
\begin{equation}
Q_A = \xi Q_B~.
\end{equation}
With this rescaling, as Brodsky emphasized,\cite{Brodsky} all the IR
renormalons are absorbed.  Obviously, it seems important to check whether
this approach really works in detail experimentally.  However, to effect
this in practice probably needs a stronger push by the theorists than
heretofore.

Another area where stronger experimentalist-theorist interactions would be of
benefit was pointed out by Zoltan Kunszt~\cite{Kunszt} in Barcelona.  Kunszt
presented a review of the recent very impressive multi-leg NLO results
obtained analytically with new techniques (helicity methods, collinear
limits, SUSY Ward identities, etc.) by Bern, Dixon and Kosower and their
collaborators.~\cite{BDK}  During the discussion of these achievements, he
bemoaned the fact that many of these results were not yet incorporated in
the experimental codes used for data analysis.  In this respect, similar
comments may well apply in the future for the heavy quark NLO results 
obtainable by the new method David Soper~\cite{Soper} discussed at
RADCOR 98.  The idea behind this method is to perform the diagramatic calculations involved 
by integrating over the energies first, deforming the relevant
contours as necessary.  In this way one is left with momentum integrals of sums
over cuts of {\bf IR insensitive} functions, allowing better numerical handling
of mass effects.  Although it remains to be seen whether this method will
prove effective, clearly if it does it also will need to be incorporated into the
experimental codes.

Let me close this section by making one more comment along this vein, but
now putting the onus on the theorists rather than on the experimentalists!
Bethke,\cite{Bethke} in his talk at RADCOR 98 also briefly reviewed the
status of $\alpha_s(M_Z^2)$, quoting a value for the world average of
\begin{equation}
\alpha_s(M_Z^2)|_{\rm world~ave} = 0.1190 \pm 0.006~.
\end{equation}
In the discussion that followed his talk, it was the view of many theorists
that the error on $\alpha_s$ quoted seemed too large, since the most reliable
analyses of $\alpha_s$ theoretically in the compilation of Bethke 
had all a smaller uncertainty than the
final error quoted.  Bethke's response was
that, as an experimentalist, he did not
feel he could, without reason, remove some of the less accurate 
$\alpha_s$ determinations from his
compilation, thus perhaps contributing to inflating the error.  I agree with
him completely.  If theorists feel that an error in $\alpha_s$ of $\delta
\alpha_s = 0.002$ is possible, then they should volunteer to vet what should
enter into the average and be prepared to weed 
out themselves dubious experimental determinations of
$\alpha_s$.  In doing so, they would be doing the field a great
service!

\section{Technical Progress}

Perhaps what impressed me the most at RADCOR 98, as a non-expert, was the amazing
technical wizardy at work.  Some of this technical firepower was directed
at extracting information on the heavy quark mass dependence of a variety of
two-loop calculations.

\subsection{Two-Loop Results}  

Expansion techniques of different sorts, to handle
these $m_Q$ effects, were discussed in a number of talks.\cite{Fleischer,Kuhn,Steinhauser,Beneke}  
The general idea is to reduce the calculations, through
the use of recursion relations, to sums of a few Master integrals $M_i$:
\begin{equation}
A = \sum_i c_iM_i~.
\end{equation}
One then evaluates the $M_i$'s in various tractable regions [e.g.
$q^2 \gg m_Q^2;~q^2\ll m_Q^2$; threshold] as Taylor series expansions.
After a conformal map from $q^2$ to 
$\omega = \left(1-\sqrt{1-q^2/4m_Q^2}\right)/\left(1+\sqrt{1-q^2/4m_Q^2}\right)$
one tries to reconstruct the function from the Taylor coefficients,
using Pad\'e expansions to improve the accuracy.  This procedure works
amazingly well, as K\"uhn~\cite{Kuhn} demonstrated by showing that one can
recover the Coulomb $1/v$ singularities from information coming from the
$q^2 \gg m_Q^2$ and $q^2 \ll m_Q^2$ regions.

These techniques are now largely automatized through the development of
dedicated computer programs.\cite{Steinhauser}  As a result, one has been
able to address a number of issues of interest, and some of these 
results were
discussed in Barcelona.  Specifically, K\"uhn~\cite{Kuhn} talked about the
calculation of $R_c$ and $R_b$ to $O(\alpha_s^2)$ and how these calculations
permit the reduction of the theoretical error in $\delta\Delta\alpha^5$ to
$\delta\Delta\alpha^5 = 0.00017$.  Steinhauser~\cite{Steinhauser} discussed
the $O(\alpha_s^2)$ corrections to $H\to t\bar t$ and the $O(\alpha\alpha_s)$
corrections to $Z\to b\bar b$ (a topic which was also discussed by Fleischer.\cite{Fleischer})  Finally, Czarnecki~\cite{Czarnecki} discussed the
$O(\alpha_s^2)$ correction to semileptonic processes, an example of which
is the process $t\to bW$ where his result is
\begin{equation}
\Gamma(t\to bW) = \Gamma_0[1-0.8\alpha_s(m_t) - 1.7\alpha_s^2(m_t)]~.
\end{equation}

One of the most interesting by-products of these technical developments is
their applicatiion to bound state problems.  Here one is interested in
examining what happens near threshold and one must develop a set of consistent
approximations which identify and include all the dominant physical
processes.  For the $b\bar b$ vacuum polarization 
graph at 2-loops, Beneke~\cite{Beneke}
integrated sequentially out different scales in the loops, 
characterized by the typical size of the loop momenta $\ell_i$ and energies
$\ell_o$.  Integrating out the hard scales $\ell_o\sim\ell_i
\sim m_b$ 
replaces the QCD Lagrangian by an effective non-relativistic Lagrangian.
Integrating out next the soft scales $\ell_o\sim \ell_i \sim m_bv$ modifies
this Lagrangian further, producing a non-local potential.  Finally,
integrating out the scales $\ell_o\sim m_bv^2$, $\ell_i\sim m_bv$ sums
up all the $1/v$ Coulomb terms, leaving just ultrasoft terms.

Beneke matched the threshold results he obtained 
in this way with those
deduced by integrating directly over the $\Upsilon$-resonances, extracting from this procedure
 a (preliminary) estimate of $m_b(m_b)$.  His result~\cite{Beneke}
\begin{equation}
m_b(m_b) = (4.37 \pm 0.08)~{\rm GeV}~,
\end{equation}
if confirmed, provides an extremely accurate determination for the $b$-quark
mass.  Certainly this result is very much more accurate
than the value for $m_b$ 
obtained
from studying the process $Z\to bb(g)$.  As Rodrigo~\cite{Rodrigo} discussed
in Barcelona, using a jet algorithm and a NLO QCD calculation of $Z\to bbg$, the
DELPHI Collaboration obtains
\begin{equation}
m_b(M_Z) = (2.65 \pm 0.25 \pm 0.34 \pm 0.27)~{\rm GeV}~,
\end{equation}
where the last error is an estimate of the theory uncertainty.

A. Czarnecki~\cite{Czarnecki} applied similar threshold expansion techniques
to obtain analytically the $O(m \alpha^6)$ contribution to positronium hyperfine
splitting.  This is a very hard calculation and a real {\it tour de force},
where the use of dimensional regularization to handle divergent terms (also used
by Beneke) was crucial to obtain the final result.  These
$O(m \alpha^6)$ terms had been calculated before, but the numerical results
obtained, by three different groups, disagreed with each other.  The result of
Czarnecki's calculation of 11.8 MHz agrees with one of these numerical
results.  However, the final result obtained for the para-ortho splitting,
including this correction, $\Delta E_{\rm p-o}$ = 203392 (1) MHz does not
quite coincide with the experimental result obtained some time ago by
Vernon Hughes and collaborators:~\cite{Hughes}  $\Delta E_{\rm p-o}|_{\rm exp}$ = 203389.10 (0.74) MHz.  Thus the hyperfine splitting of positronium has still
some remaining issues to be resolved.

In Barcelona, a further impressive analytic calculation was presented---the 
$O(\alpha^2)$ corrections to $\mu$-decay, discussed by R. G. Stuart.~\cite{Stuart}  Stuart's work uses different techniques than the ones
just described, but is no less remarkable.  Writing the $\mu$-decay lifetime as
\begin{equation}
\tau^{-1} = \frac{G_F^2m_\mu^5}{192\pi^3} [1+\Delta q]~,
\end{equation}
what Stuart discussed is the $O(\alpha^2)$ correction to $\Delta q$.  Given
that the $O(\alpha)$ correction to $\Delta$q---the famous Kinoshita, Sirlin~\cite{KS} and Berman~\cite{Berman} calculation---was computed almost 40
years ago, it is particularly nice to finally have an analytic $O(\alpha^2)$
result:~\cite{Stuart}
\begin{equation}
[\Delta q]_2 = \left(\frac{\alpha}{\pi}\right)^2 [6.701 \pm 0.002]~,
\end{equation}
where the error comes from uncertainties in the hadronic contribution.
This calculation removes altogether any theoretical uncertainty in the
Fermi constant, so that the new value for $G_F$ has an error which comes entirely
from the error on the $\mu$-lifetime itself.  Hence, now
\begin{equation}
G_F = (1.16639 \pm 0.00001)\times 10^{-5}~{\rm Gev^{-2}}~.
\end{equation}

\subsection{Electroweak Results}

There was also considerable technical progress reported in 
Barcelona in the electroweak sector.
Roughly speaking, this took place in three different areas and, to my mind,
in each area 
it involved efforts of heroic proportions.  The first of these efforts
was concerned with the full calculation of $W^+W^-$ radiative corrections, taking into
account that the $W$'s indeed decay into $f\bar f$.  Going beyond the on-shell
approximation involves a multitude of problems, as was made explicit in the
talks of Dittmaier~\cite{Dittmaier} and Behrends.~\cite{Behrends}  One of these
problems is how to preserve gauge invariance in the calculation, since using
simply a finite width for the $W$'s does not guarantee this.  What can be done,
however, is to use a complex mass everywhere (including in the definition of
$\cos^2\theta_W$!), which then allows one to respect the Ward identities throughout.~\cite{Dittmaier}  The treatment of soft photons is also particularly
tricky, as was evident in Behrends' talk.\cite{Behrends}  Fortunately, the 
principal radiation occurs off the $W$'s themselves, and not the final state
fermions.  Thus the final 
results are close to those in which the $W$'s are taken to
be on shell, and one can match-on to existing calculations rather well.

A second technical thrust in the electroweak sector, which elicited considerable
discussions at Barcelona, was the construction of improved codes for LEP 200.
M. Skrzypek~\cite{Skrzypek} discussed a new YFS Monte Carlo for $WW$ and
$ZZ$ production.  This Monte Carlo program is an extension of KORAL $W$, 
including $O(\alpha)$ electroweak corrections in the $WW$ and $ZZ$ region, and
hopes to eventually achieve a 0.5\% precision.  W. Placzek~\cite{Placzek} talked
about the work presently going on to extend the YFS Bhabha Monte Carlo program
to larger angles, again for LEP 200.  The goal here is to achieve throughout
a precision $\delta\sigma/\sigma \leq 0.25\%$.  T. Riemann~\cite{Riemann}
discussed improvements in the Z Fitter $e^+e^-\to f\bar f(\gamma)$ code to
make it effective in the LEP 200 region.  In particular, one must include
appropriate radiator cuts so as to get rid of $Z$ radiative returns.
Finally, G. Passarino~\cite{Passarino} 
discussed some of the challenges one encounters in
trying to implement a program for inclusive $q\bar q$ production
$[e^+e^-\to q\bar qX]$.  Such a program must include, for example, also
$q\bar q$'s coming from $W$-decay [i.e. $e^+e^-\to W^+W^-\to q\bar qX$].
In trying to consider the totality of effects associated with inclusive
$q\bar q$ production, the hardest problems are those connected with soft
radiation.  The goal, as Passarino explained,\cite{Passarino} is to be able to
separate effectively the result into a sum of singular and non-singular
terms, with negligible interference.  However, this can be very challenging in
practice!

A third important topic of activity centered around how to reduce the
uncertainty on the Higgs mass prediction from electroweak measurements.  The
reduction of the error band on $M_H$ (the, so called, blue-band in Fig. 3)
is predicated on being able to incorporate further theoretical refinements
in the existing radiative correction programs.  Ideally, one really wants
the full $O(\alpha^2)$ corrections to $M_W$ and $\sin^2\theta^{\rm eff}_W$, since
the dominant $O(\alpha^2 m_t^4)$ corrections are quite non-negligible.
For instance, these corrections give a contribution to the $W$-mass of
$\delta M_W\sim 15~{\rm MeV}$.
In this conference, P. Gambino~\cite{Gambino} and G. Weiglein~\cite{Weiglein}
reported on some partial results for the $O(\alpha^2)$ corrections.  Gambino
discussed the $O(\alpha^2 m_t^2)$ subleading corrections, which give mass
shifts $\delta M_W\sim 10~{\rm MeV}$.  These corrections have been incorporated
in the fitting programs.  Weiglein discussed the full {\bf fermionic}
$O(\alpha^2)$ corrections.  Because these corrections are not by themselves
under control, what Weiglein computes is really the shift in $M_W$
as one changes $M_H$ away from some reference value.  Where comparable, the
computations of Weiglein and Gambino are in very good agreement with each
other, with ``theoretical" errors at most of $O(\delta M_W\sim 1~{\rm MeV})$.

\section{Future Tests}

In Barcelona, a trio of talks~\cite{Gianotti,Miller,Blondel} were presented
detailing how future machines can contribute to precision electroweak
measurements.  Although LHC, or the proposed linear $e^+e^-$ colliders and
muon colliders are primarily discovery machines, it turns out that
asking questions of how these machines can contribute to future precision
measurements provides an important and useful challenge which helps refine
one's thinking.

For each machine, the questions one asks are in general different.  For
example, for the LHC,~\cite{Gianotti} to be able to do precision measurements,
it will be necessary to 
calibrate the leptonic
energy to $2\times 10^{-4}$; measure the jet energy to $O(1\%)$; and measure
the luminosity at the 5\% level.  If one can achieve this, then using a large statistical
sample one can hope to reduce the errors in $M_W$, $m_t$, and $M_H$ to:
\begin{equation}
\delta M_W \leq 15~{\rm MeV}~; ~~~ \delta m_t \leq 2~{\rm GeV}
\end{equation}
and, for $M_H < 500~{\rm GeV}$,
\begin{equation}
\delta M_H/M_H \sim 10^{-3}~.
\end{equation}

It was clear in Barcelona that each of the three
future accelerators discussed have
complementary physics reaches.  For instance,
the $\mu^+\mu^-$ collider is an excellent Higgs factory which, for
$M_H\sim 100~{\rm GeV}$, has a chance to determine the Higgs mass and width
to very great accuracy [$\delta M_H \sim 100~{\rm KeV}$; $\delta\Gamma_H \sim
500~{\rm KeV}$].  Furthermore, as Blondel~\cite{Blondel} explained, because the
beamstrahlung is reduced by a factor of $(m_e/m_\mu)^4$ with respect to an
$e^+e^-$ collider, one can study the threshold production properties of top
much better at a $\mu$-collider than at a linear $e^+e^-$ machine.  Blondel
estimated that one could achieve an error on the top mass of
$\delta m_t = 100~{\rm MeV}$ in a $\mu$-collider.  Furthermore, if one ran the
collider as a $Z$-factory, it may be possible to actually tune the muon spin
so that it flips at each turn!  However, there are enormous technical problems
to overcome to make a muon collider a reality.  One needs to have intense sources of muons which necessitate initial proton fluxes almost a thousand
times what has been achieved to date.  Furthermore, there is no real
solution still on how to 
effectively reduce the muon phase space through cooling, or how to 
shield the detectors from the debris arising from muon decays.  So, the
potential for physics of such a machine may never be realized.

In this respect, as Miller~\cite{Miller} discussed, $e^+e^-$ colliders are much
closer to a technical solution, both for the case of the $X$-band machine proposed
by SLAC and KEK and for the superconducting rf project (Tesla) championed by 
DESY.  Although beamstrahlung will be a significant headache, these
machines can be wonderful tools for studying relatively light Higgs
bosons and possible supersymmetric partners of existing particles.  
For example, for an
integrated luminosity of $\int Ldt \sim 50~{\rm fb}^{-1}$ at
$\sqrt{s} = 500~{\rm GeV}$, the production of a 100 GeV Higgs boson will be
rather copious, with approximately 3000 $e^+e^-\to ZH$ events and 6000
$e^+e^-\to \nu\bar\nu H$ events.  With this kind of data sample, one can
hope to study the relative branching ratios of Higgs decay into fermion-antifermion pairs (mostly $b\bar b$, but also significant 
$\tau\bar\tau$ pairs) with considerable accuracy.  Furthermore, the
availability of polarized electron beams should help to disentangle the coupling structure of light supersymmetric particles, through the study of the characteristic angular correlation structure of their decay by-products.

The situation is perhaps even more favorable, in some respects, with the LHC.
As Gianotti explained,\cite{Gianotti} perhaps the most important fact is
that the LHC is under construction and should actually begin taking data
around 2005!  Furthermore, the LHC, because of its large CM energy, can look
for supersymmetry over a wide range of masses---roughly $m_{\tilde q}, m_{\tilde g}\leq 2~{\rm TeV}$.  In addition, the LHC has also the ability to do rather
precise mesurements of the masses and couplings of supersymmetic particles,
if they exist in this mass range.  In this respect, as was emphasized in
RADCOR 98 by a number of speakers, it is important to continue to refine the
strategies of how to look for possible signals, as these in general arise
from a sequence of decays.

Another important lesson for the LHC is to carefully try to incorporate the
result of radiative effects, as these can substantially change
expectations.  This was demonstrated in Barcelona in a number of instances.
For example, Djouadi~\cite{Djouadi} showed how a light stop can depress the
cross section times branching ratio for the production and decay of a light
SUSY Higgs into two photons.  Radiative effects can also change substantially
the effective $\bar tbH^+$ coupling, thus affecting the expectations both
for top decays, if the charged Higgs is light,\cite{Guasch} or
$H^+$-decays, if the charged Higgs is heavy.\cite{Coarasa}  Spira~\cite{Spira}
discussed NLO QCD corrections for the production of squarks, gluinos and
charginos, and showed that also here these can be rather substantial,
leading to typical $K$-factors of 1.2-1.5.  Clearly one will be able to much
better judge the importance of these effects when, and if, supersymmetric
particles are found.  However, one should be aware that naive lowest order
calculations may well not suffice to properly extract the underlying
supersymmetric parameters, once a signal of supersymmetry is detected.

However, as Haber~\cite{Haber} and Pokorski~\cite{Pokorski} emphasized in
Barcelona, it may not be necessary to wait for the LHC to see new physics!
The most solid prediction of having supersymmetry at the electroweak scale
is that there should exist in the spectrum a light Higgs boson.  The light
Higgs, $h$, is the lightest of the two neutral $0^+$ bosons present in 
all supersymmetric theories.  Neglecting radiative effects, its mass is bounded
by that of the $Z$:  $m_h < M_Z$.  Radiative effects, however, can give large
mass shifts so that the above bound is only indicative.  Nevertheless, as
Weiglein~\cite{Weiglein} and Haber~\cite{Haber} discussed, these radiative mass
shifts are controlable.  Indeed, one has a 2-loop RG improved result, accurate
to $O(\alpha\alpha_s)$, which gives this shift to a few percent accuracy.  This
shift is proportional to the running top mass and depends logarithmically on
the underlying supersymmetric parameters
\begin{equation}
\Delta m_h^2 \sim G_Fm^4_t(m_t)~\ln M^2_{\rm SUSY}/m_t^2~.
\end{equation}
The resulting light Higgs mass has a maximum for large $\tan\beta$ and when the
mass of the $0^-$ neutral Higgs $m_A$ is also large.  Typically, $m_h$ is
bounded by $m_h \leq 130~{\rm GeV}$, with the bound weakening as
$\tan\beta$ decreases.  For example, as de Boer~\cite{deBoer} discussed,
$m_h\leq 105~{\rm GeV}$ for $\tan\beta = 1.6$.

The above results have important experimental implications.  Already LEP 200
has good limits on $m_h$ and $m_A$, with the combined data of all four
LEP experiments at $\sqrt{s} = 183~{\rm GeV}$ giving the bounds (for
$\tan\beta > 0.8$)~\cite{Treille}
\begin{equation}
m_h > 77~{\rm GeV}~;~~m_A > 78~{\rm GeV}~.
\end{equation}
As the LEP 200 energy is moved to $\sqrt{s} = 200~{\rm GeV}$, these limits
can be pushed up probably another 10 to 15 GeV, depending on what the total
integrated luminosity will be.  Hence, it could well be that a light
Higgs---of the type predicted by supersymmetry---may be within the reach of
LEP 200.  The Tevatron too, if it can achieve higher luminosity after its
first run with the Main Injector, has a shot at discovering this physics.
For instance, if one can achieve at the Tevatron $\int {\cal{L}}~dt\sim
(20-25)~{\rm fb}^{-1}$, one can probe for the existence of a light Higgs
up to $m_h \leq 120~{\rm GeV}$,~\cite{Treille} well within the expectation of
supersymmetric models.  Obviously, this is an important goal to try to pursue!

\section*{Acknowledgements}

I would like to thank Joan Sol\'a for the wonderful hospitality afforded to me
in Barcelona.  This work was supported in part by the Department of Energy
under Grant No. DE-FG03-91ER40662, Task C.

\end{document}